\documentclass[12pt]{article}

\usepackage[dvips]{graphicx}

\begin{document}
\def\la{\mathrel{\mathpalette\fun <}}
\def\ga{\mathrel{\mathpalette\fun >}}
\def\fun#1#2{\lower3.6pt\vbox{\baselineskip0pt\lineskip.9pt
\ialign{$\mathsurround=0pt#1\hfil##\hfil$\crcr#2\crcr\sim\crcr}}}  
\def\lrang#1{\left\langle#1\right\rangle}

\begin{center}
{\bfseries DIAGNOSTICS OF QUARK-GLUON PLASMA IN ULTRARELATIVISTIC HEAVY ION COLLISIONS
BY HARD QCD-PROCESSES~\footnote{Talk given at XV International Seminar On High Energy
Physics Problems "Relativistic Nuclear Physics and Quantum Chromodynamics", Dubna,
Russia, September 25-29, 2000}}

\vskip 3mm

I.P.Lokhtin 

\vskip 2 mm 

{\small{\it M.V.Lomonosov Moscow State University, D.V.Skobeltsyn Institute of Nuclear 
Physics}}
\end{center}

\vskip 3mm

\begin{center}
\begin{minipage}{150mm}
\centerline{\bf Abstract} 
We analyze the possibilities for studying properties of dense QCD-matter, created 
in ultrarelativistic nuclear collisions, by hard QCD-production processes, so-called 
"hard" probes -- heavy quarkonia, hard jets, high mass dimuons. Special attention is 
paid to the potential of coming heavy ion experiments on Large Hadron Collider to 
observe the rescattering and energy losses of hard partons in quark-gluon plasma. \\ 
\end{minipage}
\end{center}

\vskip 6mm

\section{Introduction}

The experimental investigation of ultrarelativistic nuclear collisions offers a unique 
possibility of studying the properties of strongly interacting matter at the high 
energy density when the hadronic matter is expected to become deconfined and a gas of 
asymptotically free quarks and gluons is formed. This is called quark-gluon plasma 
(QGP), in which the colour interactions between partons are screened owing to 
collective effects (see, for example, 
reviews~\cite{muller96,lok_rev,bass_rev}). It is assumed that the conditions for 
realizing the quark-hadron phase transition which, according to the current 
cosmological ideas, occurred during the first few microseconds of the evolution of the 
Universe, can be attained as a result of the generation of strongly excited matter in 
heavy ion collisions at accelerators. 

In recent years, a great deal of attention has been devoted to the study of "hard"
probes of QGP -- heavy quarkonia, hard hadrons and jets, high mass dimuons -- which,
not forming a part of the thermalized system, carry information about the early stages 
of the system evolution. In particular, the predicted suppression of the 
$\Psi$-resonances yield due to screening of bound state $c \bar{c}$ pair ("colour 
dipole") in plasma~\cite{satz86} or dynamical dissociation  by semi-hard deconfined 
gluons~\cite{khar94} is one of the promising signals of QGP formation. Such intriguing 
phenomenon has been observed in the most central Pb-Pb collisions in the NA50 
experiment at CERN-SPS~\cite{na50}: the "anomalously" small rate of $\Psi$-resonances, 
which is inconsistent with the conventional model of pre-resonance absorption in cold 
nuclear matter. Although the interpretation of this phenomenon as a result of the 
formation of a QGP is quite plausible, some alternative explanations (like charmonia 
rescattering on comoving hadrons) can not be still fully cancelled~\cite{vogt-psi}. For 
heavier ($b \bar{b}$) systems, a similar suppression effect in QGP is expected at 
higher temperatures than for $c \bar{c}$, which are can be reached in heavy ion 
collisions at CERN-LHC collider. 

Along with the suppression of heavy quarkonia, one of the processes of interest 
is the passage through the dense matter of jets of colour-charged partons, pairs 
of which are created at the very beginning of the collision process 
(at $\la 0.01$ fm/c) due to initial hard parton-parton scatterings. 
Such jets pass through the dense matter formed due to mini-jet production at 
larger time scales ($\sim 0.1$ fm/c), and interact strongly with the constituents in 
the medium. The inclusive cross section for hard jet 
production is not sufficient at SPS, but it increases fast with the energy of 
collided nuclei. Thus these will play important role at RHIC ($\sqrt{s} = 200 A$ GeV) 
and LHC ($\sqrt{s} = 5.5 A$ TeV for lead beams). 

The challenging arising problem here is the behaviour of "colour charge" (high-$p_T$ 
parton) in a dense QCD-matter environment: medium-induced  gluon 
radiation~\cite{ryskin,gyul94,baier,zakharov,urs99} and collisional energy losses due
to elastic rescatterings~\cite{bjork82,mrow91,lokhtin1}. Since the jet rescattering 
intensity strongly increases with temperature, formation of "hot" QGP at the initial 
temperature up to $T_0 \sim 1$ GeV at LHC~\cite{eskola94} should result in 
much larger jet energy losses as compared with the "cold" nuclear matter or hadronic 
gas. 

Although the radiative energy losses of a high energy parton have been shown to 
dominate over the collisional losses by up to an order of magnitude~\cite{gyul94}, the
angular distribution of both losses is very different~\cite{lokhtin2}. The coherent LPM 
radiation (for gluons with formation times exceeding the mean free path for 
scattering in the medium) induces a strong dependence of the jet energy on the jet cone 
size $\theta_0$~\cite{lokhtin2,baier3}. With increasing of hard parton energy the 
maximum of the angular distribution of bremsstrahlung gluons has shift towards the 
parent parton direction. This means that measuring the jet energy as a sum of the 
energies of final hadrons moving inside an angular cone with a given finite size 
$\theta_0$ will allow the bulk of the gluon radiation to belong to the jet. On the 
other hand, the collisional energy loss turns out to be practically independent on 
$\theta_0$ and emerges outside the narrow jet cone, and the relative contribution of 
collisional losses would likely become significant for jets with finite cone 
size~\cite{lokhtin2}. 

\section{Hard jet production at LHC} 

It is well established now, that A Large Ion Collider Experiment (ALICE)~\cite{alice} 
will be specially dedicated to the heavy ion physics at LHC. Besides, the general 
purpose detector Compact Muon Solenoid (CMS)~\cite{cms}, which is optimized mainly for 
the search of the Higgs boson in proton-proton collisions, will be able to perform the 
accurate measurement of the characteristics of high-energy muons, photons, electrons, 
and hadronic jets, making it useful also for studying hard probes of the QGP in heavy 
ion collisions~\cite{note00-060}. 

Let us consider an example, which demonstrates the sensitivity of jet production
pattern to the medium-induced parton rescattering and energy losses:  
distribution of high-$p_T$ jet pair over the impact parameter $b$ of nucleus-nucleus 
collision~\cite{lokhtin3}. The intriguing prediction associated with the coherence 
pattern of the medium-induced radiation is that radiative energy losses per unit 
distance $dE/dx$ depend on the total distance traversed $L$, asymptotically 
$dE/dx \propto L$~\cite{baier}. Experimentally, the dependence of energy losses of the 
distance traversed can be studied in different bins of impact parameter $b$: 
the average effective transverse size of dense zone $\left< L \right>$
decreases almost linearly with increasing $b$~\cite{lokhtin3}. 

The details of the geometrical model for jet production and jet passing through a dense 
matter, created in nuclear overlapping zone, can be found in our recent 
work~\cite{lokhtin3}. In general, the intensity of rescattering and energy losses are 
sensitive to the initial parameters (energy density $\varepsilon_0$, formation time 
$\tau_0$) and space-time evolution of a matter. We treat here the medium as a 
longitudinally expanding quark-gluon fluid, and partons as being produced on a 
hyper-surface of equal proper times $\tau = \sqrt{t^2 -  z^2}$~\cite{bjorken}. The 
total energy losses in transverse direction are the result of averaging over the whole 
"nuclear geometry" (i.e. jet production vertices and azimuthal angles) at given 
$b$~\cite{lokhtin3}:   
\begin{equation}
\label{en_los} 
\left< \Delta E_T (b)\right> = \left<  
\int\limits_{\displaystyle\tau_0}^{\displaystyle 
\tau_L}d\tau \left( \frac{dE}{dx}^{rad}(\tau) + \sum_{b}\sigma_{ab}(\tau)\cdot
\rho_b(\tau)\cdot \nu(\tau) \right) \right> _{"nuclear~geometry"}.    
\end{equation} 
Here $\tau_0$ and $\tau_L$ are the proper time of the plasma formation and 
the time of jet escaping from the dense zone respectively; $\rho_b \propto T^3$ is the 
density of plasma constituents of type $b$ at temperature $T$; $\sigma_{ab}$ is the 
integral cross section of scattering of the jet parton $a$ off the comoving constituent 
$b$ (with the same longitudinal rapidity $y$); $\nu$ and $dE / dx ^{rad}$ are 
the thermal-averaged collisional energy loss of a jet parton due to single elastic 
scattering and radiative energy losses per unit distance respectively. The formulas for 
collisional and radiative losses can be found in~\cite{lokhtin3} (see also references 
therein). Let us just to mark the following essential difference between collisional 
and radiative losses behaviour. 

The collisional loss per scattering $\nu$ is independent of total distance traversed 
and determined by temperature, roughly $\nu \propto T$. Then total collisional energy 
losses integrated over whole jet path are estimated as $\left< \Delta E_{col} \right> 
\propto T_0^2 \propto \sqrt{\varepsilon_0}$. The $\tau_L$-dependence of 
$\Delta E_{col}$ can be weaker than linear for expanding medium ($\Delta E_{col} 
\propto \tau_L$ for static matter). On the other hand, for medium-induced coherent 
gluon radiation in the limit of "strong" LPM effect, $\omega \gg \mu_D^2\lambda_g$ 
($\omega$, $\mu_D$ and $\lambda_g$ are the energy of radiated gluon, Debye screening
mass and gluon mean free path in the medium respectively), we 
have~\cite{baier,baier3} ${dE}/{dx}^{rad} \propto T^3$ and ${dE}/{dx}^{rad} \propto 
\tau_L$ with logarithmic accuracy. Then total radiative energy losses $\left< \Delta 
E_{rad} \right> = \int d\tau \cdot {dE}/{dx}^{rad}$ are estimated as $\left< \Delta 
E_{rad} \right> \propto T_0^3 \propto \varepsilon_0^{3/4}$ and $\Delta E_{rad} \propto 
\tau_L^{\beta}$ ($\beta \la 2$). 

In our calculations we used the well-known scaling Bjorken's solution~\cite{bjorken} 
for QGP density: 
$~\varepsilon(\tau) \tau^{4/3} = \varepsilon_0 \tau_0^{4/3},~ 
T(\tau) \tau^{1/3} = T_0 \tau_0^{1/3},~ \rho(\tau) \tau = \rho_0 \tau_0$. 
For certainty we used the initial conditions for the gluon-dominated plasma  
($N_f \approx 0$, $\rho_q \approx 1.95T^3$) expected for central $Pb-Pb$ collisions at 
LHC~\cite{eskola94}: $\tau_0 \simeq 0.1$ fm/c, $T_0 \simeq 1$ GeV. Note 
that initial energy density $\varepsilon_0$ in dense zone depends on $b$ very slightly 
($\delta \varepsilon_0 \la 10 \%$) up to $b \sim R_A$~\cite{lokhtin3}. Thus the 
variation of impact parameter $b$ of nuclear collision (which can be measured 
using the total transverse energy deposition in calorimeters) up to $b \sim 
R_A$ allows one to study jet quenching as a function of distance traversed 
without sufficient changing $\varepsilon_0$.   

The rate of $\{ij\}$ type dijets with transverse momenta $p_{T1}, p_{T2}$ 
produced in initial hard scatterings in $AA$ collisions at given $b$ is 
obtained by averaging over "nuclear geometry": 
\begin{equation} 
\frac{dN_{ij}^{dijet}}{dp_{T1}dp_{T2}d^2b} = \frac{1}{\sigma^0_{jet}}\frac{d^2
\sigma^0_{jet}}{d^2b}\cdot \left< \int dp_T^2\frac{d\sigma_{ij}} {dp_T^2}~\delta(p_{T1}
- p_T + \Delta E_T^i)~ \delta(p_{T2} - p_T + \Delta E_T^j) \right> ,
\end{equation} 
where $d\sigma_{ij} / dp_T^2$ is the parton differential pQCD cross section. The 
integrated above threshold $p_T^{cut}$ dijet rate in $AA$ relative to $pp$ 
collisions can be studied at LHC with a reference process, unaffected by energy loss  
and with a production cross section proportional to the number of nucleon-nucleon 
collisions, such as Z$(\rightarrow \mu^+\mu^-)$ production~\cite{note00-060}.   
The cross section $d^2 \sigma^0_{jet} / d^2b$ for initially produced jets in 
$AA$ collisions at given $b$ is~\cite{note00-060}: 
\begin{equation} 
\label{jet_prob}
\frac{d^2 \sigma^0_{jet}}{d^2b} ({\bf b}, \sqrt{s}) =  T_{AA} ({\bf b}) 
 \sigma _{NN}^{jet} (\sqrt{s}) \left[ 1 - \left( 1- \frac{1}  
{A^2}T_{AA}({\bf b}) \sigma^{in}_{NN} (\sqrt{s}) \right) ^{A^2} \right] 
\end{equation} 
where $T_{AA}(b)$ is the nuclear overlap function; $\sigma _{NN}^{jet}$ and 
$\sigma^{in}_{NN}$ are the computed with PYTHIA$5\_7$ model~\cite{pythia}  
$NN$ hard scattering cross section and total inelastic cross section respectively.  

\begin{figure}[hbtp]
\centerline{
\includegraphics[height=77mm]{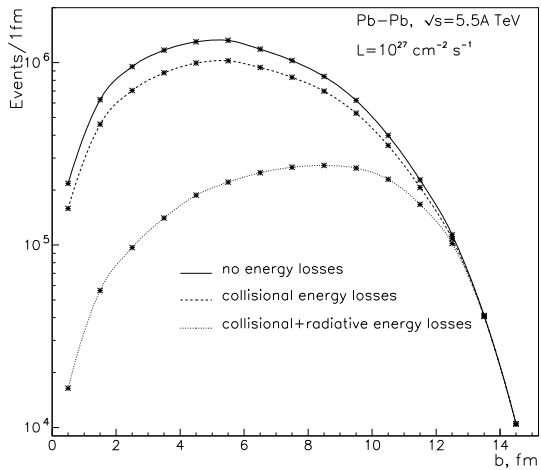}}
\caption{The distribution of dijet rates over impact parameter: no energy losses (solid 
curve), with collisional losses (dashed), with collisional and radiative losses 
(dotted).}
\end{figure}

Figure 1 shows dijet rates ($E_T^{jet}> 100$ GeV, CMS rapidity acceptance $|y^{jet}| 
< 2.5$) in different impact parameter bins for three cases: $(1)$ without energy 
losses, $(2)$ with collisional losses only ($\theta_0 \gg 0$), $(3)$ with collisional 
and radiative losses ($\theta_0 \rightarrow 0$). The rates are normalized to the 
expected number of events produced in Pb-Pb collisions during two weeks ($1.2 \times 
10^6$ s) of LHC run time, assuming luminosity $L = 10^{27}~cm^{-2}s^{-1}$~\cite{cms}. 
The total initial dijet rate with $E_T^{jet} > 100$ GeV is estimated as $1.1 \times 
10^{7}$ events. Since the jet quenching is much stronger in central collisions than in 
peripheral one's, the maximum and mean values of $dN^{dijet}/db$ distribution get 
shifted towards the larger $b$ as compared to what is expected from independent 
nucleon-nucleon interactions pattern. 

\section{High mass dimuon production at LHC} 

Let us consider now the high invariant mass ($M_{\mu^+\mu^-} \ga 20$ GeV/c$^2$) dimuon 
production. One of the main dimuon sources in this "resonances-free" mass region is the 
semileptonic $B$-meson decays, which are expected to be sensitive to the rescattering 
of massive $b$ quarks in dense medium~\cite{dimuons}. The main correlated background, 
Drell-Yan dimuons, can be rejected using the tracker information on secondary vertex 
position. The uncorrelated part of dimuon background (random decays of pions and kaons, 
muon pairs of mixed origin) can be subtracted using like-sign dimuon mass spectra
samples~\cite{note00-060}. 

The correct description of coherent radiation experienced by massive quark is still not 
solved task. On the one hand, the medium-induced radiation of slow quarks, $p_T \la 
M_q$, should be suppressed by their mass. On the other hand, the ultrarelativistic 
limit, $p_T \gg M_q$, corresponds to the radiation spectrum of massless quark. In our 
case, the main contribution in the mass region $M_{\mu^+\mu^-} = 20\div 50$ GeV/$c^2$ 
is due to b-quarks with "intermediate" values of $p_T \ga 5\div 10$ GeV/$c$. We
estimated the sensitivity of dimuon spectra to the medium-induced effects for two 
extremal cases: $(1)$ collisional energy losses only (lower limit); $(2)$
collisional and radiative energy losses without taking into account LPM coherent 
suppression of radiation, i.e. $dE/dx \propto E$ and does not depend on $L$ here 
(upper limit). We performed the Monte-Carlo simulation of mean free path of $b$-quarks 
of mass $M_b = 5$ GeV/$c^2$ in QGP, formed in nuclear overlapping zone 
(see the previous chapter). The nuclear shadowing (which is $\sim 15\%$ 
in this kinematic region) and transverse momentum kicks obtained by $b$-quark per 
rescattering have been taken into account. 

\begin{figure}[hbtp]
\centerline{
\includegraphics[height=77mm]{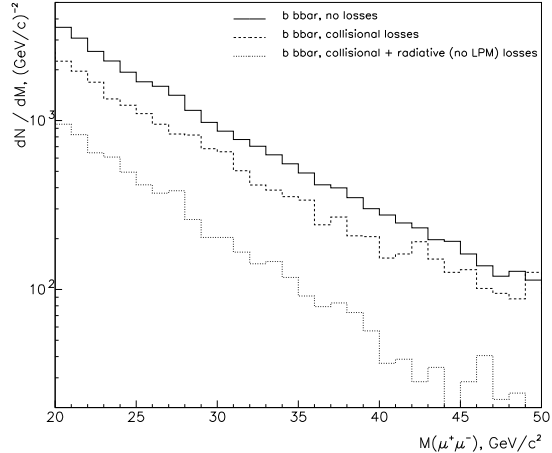}}
\caption{Invariant mass spectrum of $\mu^+\mu^-$ pairs, $p_T^{\mu} > 5$ GeV/c: 
no energy losses (solid histogram), with collisional losses 
(dashed), with collisional and radiative losses (dotted).}
\end{figure}

Figure 2 presents the  $\mu ^+ \mu^-$ invariant mass spectra from $b$-decays without 
and with energy losses of $b$-quarks at CMS muon acceptance: 
$p_T^{\mu} > 5$ GeV/$c$, $|\eta^{\mu}| < 2.4$. The rates are normalized again to the 
number of events produced in Pb-Pb collisions for two weeks of LHC run 
time. The total initial dimuon rate at $M_{\mu^+\mu^-} = 20\div 50$ GeV/$c^2$ is 
estimated with PYTHIA$5\_7$ and CTEQ2L parton distribution function as $2.8 
\times 10^{4}$, and it can be reduced by factor $\sim 1.6\div 4$ due to rescattering 
and energy losses of $b$-quarks in QGP. The absolute rates here, of course, depend on 
the PDF choice, $B$-meson fragmentation scheme, next-to-leading order corrections, 
etc. Thus the high mass dimuon spectra measurements with proton or deuterium beams at 
the same energy per nucleon are requested. 

\section{Conclusion}
The recent data obtained at CERN-SPS with heavy ion beams (charmonium suppression, 
strangeness enhancement) are consistent with the predicted signatures of a quark-gluon 
plasma and can not be explained using "conventional" hadronic interactions scenarios. 
In order to study the properties of the deconfined state in the coming heavy 
ion experiments at BNL-RHIC and CERN-LHC, the "hard probes", as heavy quarkonia and 
high-$p_T$ jets, can be certainly useful. Since such objects are created at the very 
beginning of the nuclear collision process, these propagates through the dense matter 
formed at larger time scales, and interact strongly with constituents in the medium. We 
have shown that distribution of jets over impact parameter and high-mass dimuon spectra 
in heavy ion collisions at LHC are quite sensitive to the medium-induced parton 
rescattering and energy losses.

\end{document}